\documentclass[usenatbib,useAMS]{mn2e}
\usepackage{epsfig}
\usepackage{amsmath,amssymb,bm}
\usepackage{color}
\usepackage[varg]{txfonts}
\usepackage{aas_macros}
\usepackage{pdfpages}

\title[Formation of Kozai--Lidov Planets]{Planet-disc evolution and the formation of
  Kozai--Lidov planets}

\author[Martin et al.]{Rebecca G. Martin$^{1}$, Stephen H. Lubow$^2$,
  Chris Nixon$^3$ and Philip J. Armitage$^{4,5}$\\ $^{1}$Department of
  Physics and Astronomy, University of Nevada, Las Vegas, 4505 South
  Maryland Parkway, Las Vegas, NV 89154, USA \\ $^2$Space Telescope
  Science Institute, 3700 San Martin Drive, Baltimore, MD 21218,
  USA\\ 
  $^3$Department of Physics \& Astronomy, University of Leicester, Leicester, LEI 7RH, UK \\
  $^4$JILA, University of Colorado \& NIST, UCB 440, Boulder, CO
  80309, USA \\
  $^5$Department of Astrophysical \& Planetary Sciences, University of Colorado,
  Boulder, CO 80309, USA}

\date{}
\pagerange{\pageref{firstpage}--\pageref{lastpage}} 
\pubyear{2016}
 
\begin{document}
\maketitle
\label{firstpage}

\begin{abstract}

With hydrodynamical simulations we determine the conditions under
which an initially coplanar planet--disc system that orbits a member 
of a misaligned binary star evolves to form a planet that undergoes
Kozai--Lidov (KL) oscillations once the disc disperses. These
oscillations may explain the large orbital eccentricities, as well as
the large misalignments with respect to the spin of the central star,
observed for some exoplanets.  The planet is assumed to be massive enough to open a
gap in the disc. The planet's tilt relative to the binary orbital
plane is subject to two types of oscillations. The first type, present
at even small inclination angles relative to the binary orbital plane,
is due to the interaction of the planet with the disc and binary
companion and is amplified by a secular resonance.  The second type of
oscillation is the KL oscillation that operates on both the
planet and disc at larger binary inclination angles.  We find that for
a sufficiently massive disc, even a relatively low inclination
planet--disc system can force a planet to an inclination above the
critical KL angle, as a consequence of the first type of tilt
oscillation, allowing it to undergo the second type of oscillation.
We conclude that the hydrodynamical evolution of a sufficiently
massive and inclined disc in a binary system broadens the range of
systems that form eccentric and misaligned giant planets to include a
wide range of initial misalignment angles ($20^\circ\lesssim
i\lesssim160^\circ$).

\end{abstract}


\section{Introduction}
\label{intro}
To date, a total of 1642 exoplanets have been confirmed \citep[see  exoplanets.org,][]{Wrightetal2011,Han2014}, of 
which an estimated $40-50\%$ are in binary systems \citep{Horchetal2014}.  Thus,
it is important to understand planet formation and evolution in binary
systems in order to explain the observed exoplanet orbital and
physical properties. In particular, we would like to quantify the role of 
binaries in the formation of eccentric and inclined planetary systems. 
At radii where tidal effects may be neglected, most massive extrasolar 
planets have significantly eccentric orbits, with some planets having extreme 
eccentricities $e>0.9$ \citep{Tamuz2008,OTool2009,Moutou2009}. For 
(typically) smaller separation systems, where the misalignment of the planetary orbit 
to the spin of the host star can be measured with the Rossiter--McLaughlin effect
\citep{Rossiter1924,McLaughlin1924,Queloz2000}, misalignments are common 
\citep[e.g.][]{Triaudetal2010,Schlaufman2010,Winn2010}\footnote{See the 
online compilation by R. Heller http://www.physics.mcmaster.ca/$\sim$rheller/}. 
Some planets even have retrograde orbits \citep[e.g][]{Winn2009,Lund2014}. 
The observations are broadly consistent with a model in which the obliquities 
of close-in giant planets were initially randomly distributed, with alignment 
subsequently occurring in a subset of systems where the tidal timescale 
is short \citep{Albrechtetal2012}.

Very massive planets (or brown dwarfs) that open broad gaps in the protoplanetary 
disc can be excited to significant eccentricities by gravitational planet--disc 
interactions \citep{Papaloizouetal2001,Dunhill2013}. For roughly Jupiter mass planets, 
excitation and damping processes are in near balance 
\citep{Ogilvie2003,Goldreich2003}, and whether there is any net excitation 
of eccentricity may depend upon details of the disc structure \citep{Tsang2014}. 
Simulations, however, suggest that even when eccentricity can be excited the 
resulting eccentricities are relatively small \citep{DAngelo2006,Duffell2015}, 
and moreover inclinations with respect to the disc plane are more securely in the damping regime 
\citep{XiangGruess2013,Bitsch2013}. Hence, the subset of planets with highly eccentric 
or significantly misaligned orbits probably suffered dynamical excitation of their orbits, 
through a combination of planet--planet scattering and secular perturbations of various 
kinds \citep{FordRasio2008,Chatterjee2008,Dawson2013,Petrovich2014}. In this work we focus 
on a subset of secular perturbations that could produce the very high eccentricity planets. 

The Kozai--Lidov \citep[KL,][]{Kozai1962,Lidov1962} mechanism is a secular effect that can form 
highly eccentric and inclined exoplanets \citep[e.g.][]{Wu2003,Takeda2005,Wuetal2007,Nagasawa2008,Perets2009}. 
In the simplest case where the perturber is a companion in binary star system, the 
process periodically exchanges the inclination and eccentricity of a test particle in a misaligned orbit around one component of the binary if the initial inclination, $i_{0}$, satisfies $\cos^2
i_{0}<\cos^2 i_{\rm crit}=3/5$ \citep[e.g.][]{Innanen1997}. The component of angular momentum that is perpendicular to the binary orbital plane, as well as energy, is nearly conserved during this process so that
\begin{equation}
\sqrt{1-e^2}\cos i \approx \rm const.
\end{equation}
As the eccentricity of the particle increases from $0$ up to
\begin{equation}
e_{\rm max}=\sqrt{1-\frac{5}{3}\cos^2 i_0},
\end{equation}
its inclination decreases from $i_0$ to $i_{\rm crit}$ and vice versa. The oscillations repeat exactly in time for the test particle case in a circular binary (when the test particle is close to its host star). However, if the binary orbit is eccentric, test particle orbits can instead evolve to a retrograde orientation with chaotic behaviour \citep[e.g.][]{Naoz2011,Lithwick2011,Li2014}. For the observed highly inclined planets, there may be some later circularization due to tidal friction after the eccentricity growth \citep*{Fabrycky2007, Naozetal2012}. Thus, the planets can end up on a circular but highly inclined orbit.

In order to understand planet formation in binaries, we must first understand the birthplace, the protoplanetary disc. In a young binary star system, the protoplanetary disc around each star may be misaligned with respect to the binary orbital plane.  There is some direct observational evidence of this in wide binary systems \citep[e.g.][]{Jensenetal2004,Skemer2008,Roccatagliata2011}. For example, the young binary system HK Tau with a projected separation of about $350 \,\rm AU$, has discs observed around both components with one disc edge-on and the other closer to face-on \citep{Stapelfeldt1998}. Although the binary orbit is unknown, at least one of discs must be substantially misaligned with the binary orbital plane. ALMA observations suggest that the misalignment between the two discs is $60^\circ- 68^\circ$ \citep{Jensen2014}.  Furthermore, a wide binary, with a projected separation of about $440 \,\rm AU$, in Orion has a misalignment between the projected disc rotation axes of about $72^\circ$ \citep{Williams2014}. Misaligned discs in binaries suggest that wide binary star systems do not form directly from a single large corotating primordial structure but rather they are subject to small scale effects, such as turbulence, that cause a lack of correlation between the rotational axes of the accreting gas on to each star \citep{Offner2010,Tokuda2014,Bateetal2010,Bate2012}. Thus, misaligned discs could be typical rather than special cases.

\begin{figure}
\centering
\includegraphics[width=6.0cm]{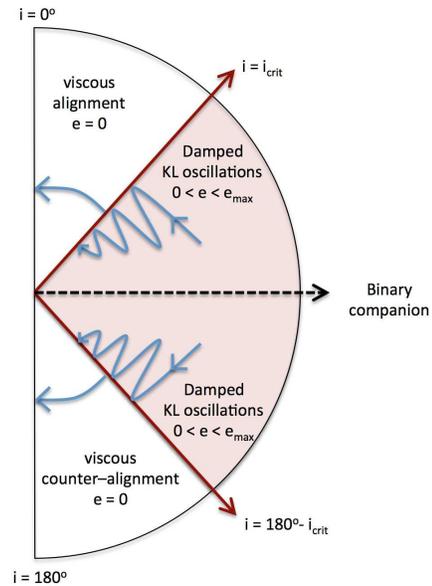}
\caption{The alignment process of a misaligned disc around one
  component of a binary system.  The inclination angle $i$ is 
   measured between the angular momentum vector of the disc and the
    angular momentum vector of the binary orbit. At an inclination of zero,
    the disc lies in
    the binary orbital plane. At inclination of $180^\circ$, the disc also in
    the binary orbital plane but rotates in a retrograde
    sense.   If
  the disc inclination lies in the range $i_{\rm
    crit}<i<180^\circ-i_{\rm crit}$, the disc displays damped KL
  oscillations where the inclination and eccentricity of the disc are
  exchanged. Because these oscillations are damped, the
    inclination tends towards the critical KL angle.  If the
    inclination is below the critical KL angle, the alignment
  proceeds monotonically towards alignment or counter--alignment with
  zero eccentricity. Note that this assumes that self--gravity is not
  strong enough to prevent the KL oscillations.}
\label{klalign}
\end{figure}

Recently, we found that a highly misaligned fluid disc around one
component of a binary system may also be unstable to global KL
oscillations \citep{Martinetal2014b, Fu2015}. That is, oscillations
occur where the disc inclination and eccentricity are periodically
exchanged. In a test particle situation, the KL oscillations continue
without damping. However, because of dissipation within the fluid
disc, the oscillations damp until the disc becomes circular and
inclined at roughly the critical angle for KL oscillations to occur,
$i_{\rm crit}$. The disc continues to align with the binary orbital
plane on a longer timescale by viscous torques that interact with a
disc warp \citep[e.g.][]{Kingetal2013}. Fig.~\ref{klalign} sketches
the alignment process of the disc with the binary orbital plane. If
the disc is closer to counter--alignment than alignment, it moves
towards counter--alignment. The complete realignment process may take
longer than the disc lifetime.

At early times in the disc evolution, the disc mass is large and the
disc may be self-gravitating. Self-gravity may be sufficiently strong
to suppress the KL oscillations for some period of time
\citep{Batygin2011, Batygin2012,Fu2015b}.  Instead, the disc globally
precesses as a rigid body, while its inclination gradually decreases
monotonically (as is the case for disc inclinations below the critical
KL angle). As material drains on to the central star, eventually
self-gravity weakens and the KL cycles begin if the disc is still
sufficiently misaligned.  For a highly inclined
disc of planetesimals, the KL mechanism strongly inhibits
planetesimal accumulation and thus planet formation is difficult. 
Thus, in the core accretion model, planets in
highly misaligned systems may need to form either before the KL
oscillations, or after they have finished as planet formation may be
difficult while the KL effect is in operation
\citep[e.g.][]{Marzarietal2009,Fragneretal2011}. 

In a recent paper, we investigated the evolution of a giant planet in
a slightly misaligned disc around one component of a binary star
system \citep{Lubow2015b}. Initially, the orbit of the planet
and the disc are taken to be coplanar. We considered initial tilt angles relative
to the binary orbital plane to be well below the
critical KL angle. We found that as the system evolves, the planet 
orbit is generally not coplanar with the disc.
It only remains coplanar to
the disc in the limit of a low mass planet that does not open a substantial gap. 
The relative inclination of the planet--disc system undergoes secular oscillations 
whose amplitudes are generally of order the initial tilt of the system relative to
the binary orbital plane. 
For disc masses that are somewhat large compared the planet mass, the disc and planet precess together, while both undergo oscillations in their tilts relative
to the orbital plane of the binary.
 This effect
can cause the planet's orbital tilt to increase above its initial value. Such tilt changes
may have consequences
on the evolution of an initially low inclination planet into KL oscillations that we explore here.

If a planet remains at or below the critical KL angle after the disc dispersal, the planet remains stable against KL oscillations, unless it is perturbed (by another planet, for example).  In this work, we focus on the evolution of more highly misaligned protoplanetary discs ($20^\circ<i<60^\circ$) that contain a giant planet around one component of a binary star system.  We consider only the case where the planet is large enough to open a gap in the protoplanetary disc. Very recently, \cite{Picogna2015} considered such evolution in the context of planet migration through tilted discs in binaries. Our simulations span a much wider range of parameter space,  including smaller initial tilt angles relative to the binary orbital plane. In Section~\ref{sph} we use three dimensional hydrodynamical simulations to investigate the evolution and parameters that allow the formation of a KL planet.  In Section~\ref{discussion} we discuss the stages of formation of a KL planet. We present our conclusions in Section~\ref{conc}.

\section{Hydrodynamical Planet--Disc--Binary Simulations}
\label{sph}

In this Section we consider the evolution of a protoplanetary disc that contains a planet, around one component of a circular orbit, equal mass,  misaligned binary star system.  We use the smoothed particle hydrodynamics (SPH; e.g. \citealt{Price2012a}) code {\sc phantom} \citep{PF2010,LP2010}.  {\sc Phantom} has been used extensively for modelling discs with complex geometries around single stars and binary systems \citep[e.g.][]{Nixon2013,Martinetal2014a,Dogan2015,Nealon2015,Nealon2016}.  The binary star system, disc, and planet parameters are summarised in Table~\ref{tab}.  The equal mass binary star, with total mass $M=M_1+M_2$, has a circular orbit in the $x$-$y$ plane with separation, $a$. The mass of the planet is $M_{\rm p}=0.001\,M$ and its initial separation to the primary star is $0.1\,a$. We choose the accretion radius of each star to be $0.025\,a$ and the planet to be $0.005\,a$. The stars and the planet are treated as sink particles. Disc material that falls within their accretion radius is added to the sink while conserving mass, and momentum. The planet and the disc begin coplanar to each other but misaligned to the binary orbital plane. The  two parameters that we change between our simulations are the mass of the disc and the misalignment angle with respect to the binary. 

\begin{table*}
\caption{Parameters of the initial disc set up for a circular equal
  mass binary with total mass, $M$, and separation, $a$.} \centering
\begin{tabular}{lllll}
\hline
Simulation Parameter & Symbol & Value(s) \\
\hline
\hline
Mass of binary component &  $M_1/M = M_2/M$ & 0.5 \\
Accretion radius of the binary masses & $R_{\rm acc}/a$    & 0.025  \\
\hline
Initial disc mass & $M_{\rm di}/M$ & [0.001 ,0.008, 0.01] \\
Initial disc inner radius & $R_{\rm in}/a$ & 0.025 \\
Initial disc outer radius & $R_{\rm out}/a$ & 0.25 \\
Disc viscosity parameter & $\alpha$ & $0.05$ \\
Disc aspect ratio    & $H/R (R=R_{\rm in})$ & 0.036 \\
Initial disc inclination & $i/^\circ$ & [20, 40, 60] \\ 
\hline
Planet Mass & $M_{\rm p}/M$ &  0.001 \\
Initial planet inclination & $i_{\rm p}/^\circ$ &[20, 40, 60] \\
Initial planet separation to primary & $a_{\rm p}/a$ & 0.1 \\
Accretion radius of the planet & $R_{\rm p,acc}/a$ & 0.005 \\
\hline
\end{tabular}
\label{tab}
\end{table*}

We follow the methods of \cite{Lubow2015b} for the initial disc setup. We simulate a low mass disc ($M_{\rm d}=10^{-6}M$) that is coplanar to the binary and the planet in a simulation for 10 binary orbits so that the surface density profile has the correct gap size around the planet.   The disc has $500,000$ SPH particles. The surface density of the disc is initially distributed as a power law in radius $\Sigma \propto R^{-3/2}$ between $R_{\rm in}=0.025\, a$ and $R_{\rm out}=0.25\,a$. The outer radius of the disc is the tidal truncation radius for a coplanar disc \citep{Paczynski1977}. The disc is locally isothermal with sound speed $c_{\rm s} \propto R^{-3/4}$ and $H/R=0.036$ at $R=R_{\rm in}$. These parameters allow $\alpha$ and $\left<h\right>/H$ to be constant over the disc \citep{LP2007}. The \cite{SS1973} $\alpha$ parameter is taken to be 0.05 (we implement the disc viscosity in the usual manner by adapting the SPH artificial viscosity according to the procedure described in \cite{LP2010}, using $\alpha_{\rm AV} = 0.95$ and $\beta_{\rm AV} = 2.0$). The disc is resolved with shell-averaged smoothing length per scale height $\left<h\right> /H \approx 0.52$.

We do not include the effects of self--gravity in our simulations. In
\cite{Fu2015b}, we found that self--gravity may suppress the KL
oscillations of a disc, but only in relatively high mass discs,
$M_{\rm d}\gtrsim 0.01\,M$. Here, we consider only disc masses in the
range $0.001\le M_{\rm d}/M\le 0.01$. Thus, the simulations presented
in this work would not be significantly affected by the inclusion of
self--gravity. However, for discs with a higher mass, self--gravity
may play some role in the evolution.

\begin{figure}
\begin{center}
\includegraphics[width=6.5cm]{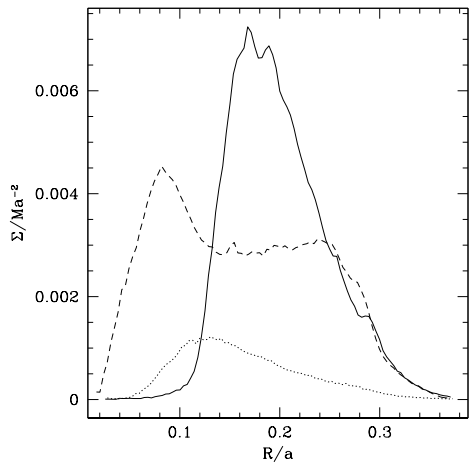}
\end{center}
\caption{Surface density distribution for the disc model with initial
  misalignment $i=60^\circ$, disc mass $M_{\rm d}=0.001\,M$, where $M$ is the mass of the binary. The solid
  line shows the initial distribution at $t=0$ (after an initial stage that begins with a coplanar system that has no disc gap), the dashed line shows
  $t=6\,P_{\rm b}$ and the dotted line $t=40\,P_{\rm b}$.}
\label{surfdens}
\end{figure}

During the initial 10 binary orbital periods that we ran the
simulation in a coplanar sense, the planet carved out a gap in the
disc. The gap restricts the communication between the inner and outer
parts of the disc and thus they behave independently. During this
time, the majority of the material interior to the planet is
accreted on to the primary star, and some on to the planet. But since the disc mass is very small, the increase
in the planet mass is also very small. After 10 binary orbits, we then tilt the disc and planet orbit (such that they remain coplanar) and increase the mass of the disc to the required amount. The solid line in Fig.~\ref{surfdens} shows the initial surface density profile for a disc with mass $M_{\rm d}=0.001\,M$. In a misaligned disc, the binary torque is weaker and thus the disc can spread outwards during the initial evolution \citep{Lubowetal2015}.  A misaligned disc and planet undergo retrograde nodal precession due to the torques caused by the companion star. That is, the orbital axis of the disc precesses about the axis of rotation of the binary. For typical protostellar disc parameters, the disc remains nearly flat and undergoes little warping as it precesses
\citep{Larwoodetal1996}. Typically the precession occurs on a shorter timescale than the alignment timescale.

In \cite{Lubow2015b} we studied the evolution of such a system with a
small misalignment angle ($i = 10^\circ$), well below the critical KL
angle.  In Section~\ref{hi} we first consider the evolution for high
misalignment angles, followed by critically misaligned systems in
Section~\ref{crit} and then finally some low inclination cases which
are still able to form a planet that exhibits KL cycles in
Section~\ref{low}.

\subsection{Highly Misaligned Systems}
\label{hi}   

\begin{figure*}
\begin{center}
\includegraphics[width=6.5cm]{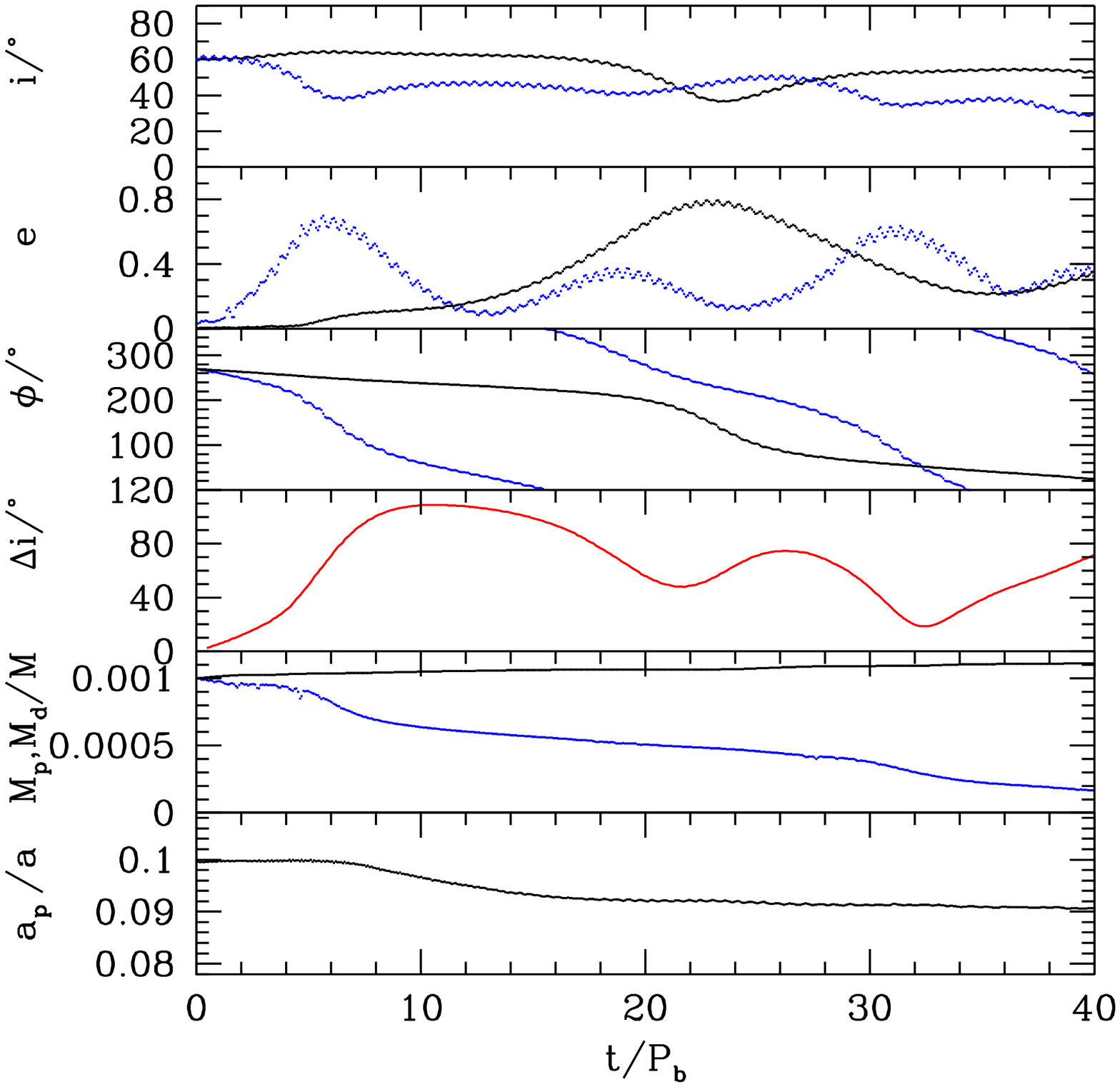}
\includegraphics[width=6.5cm]{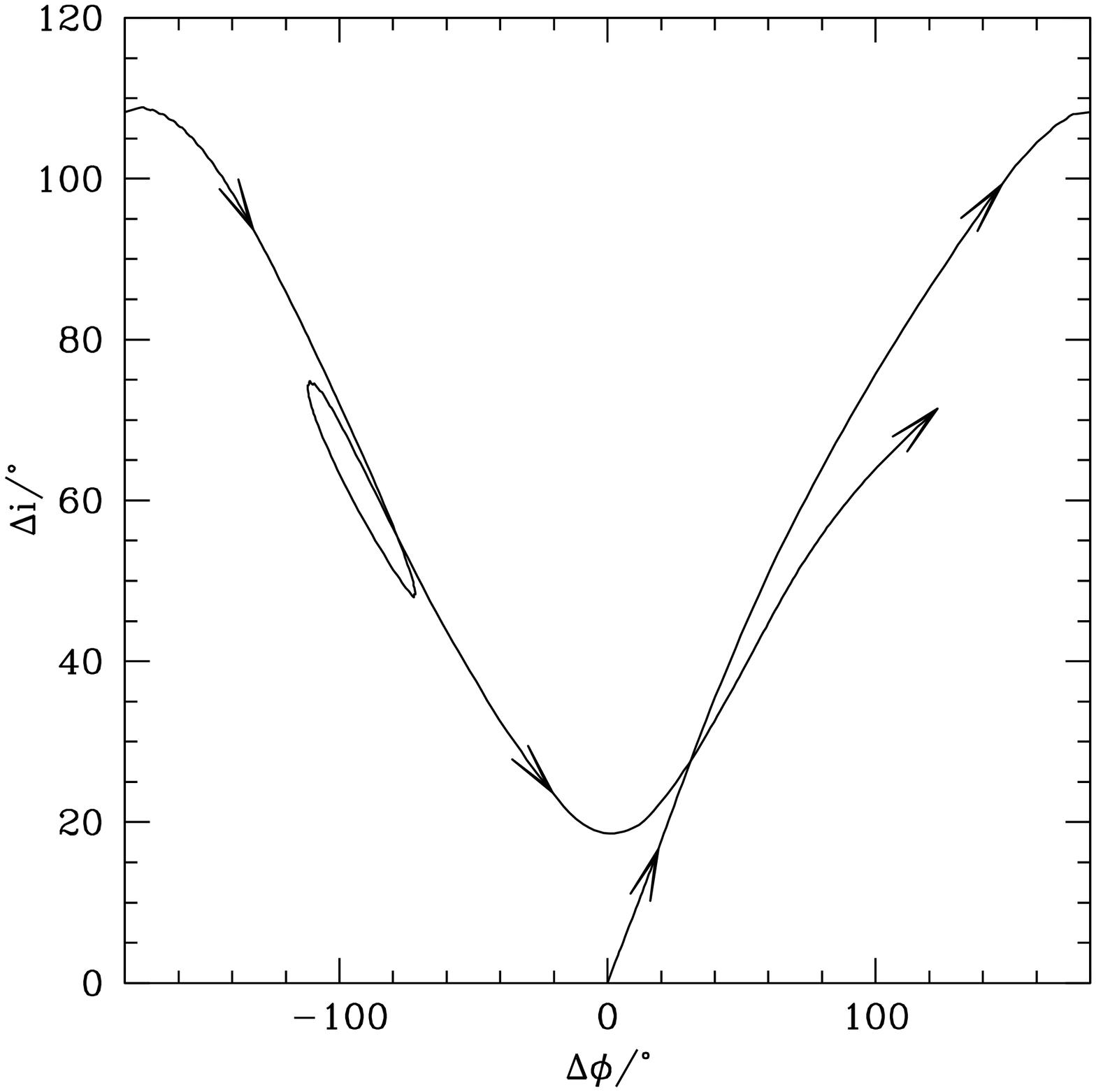}
\includegraphics[width=6.5cm]{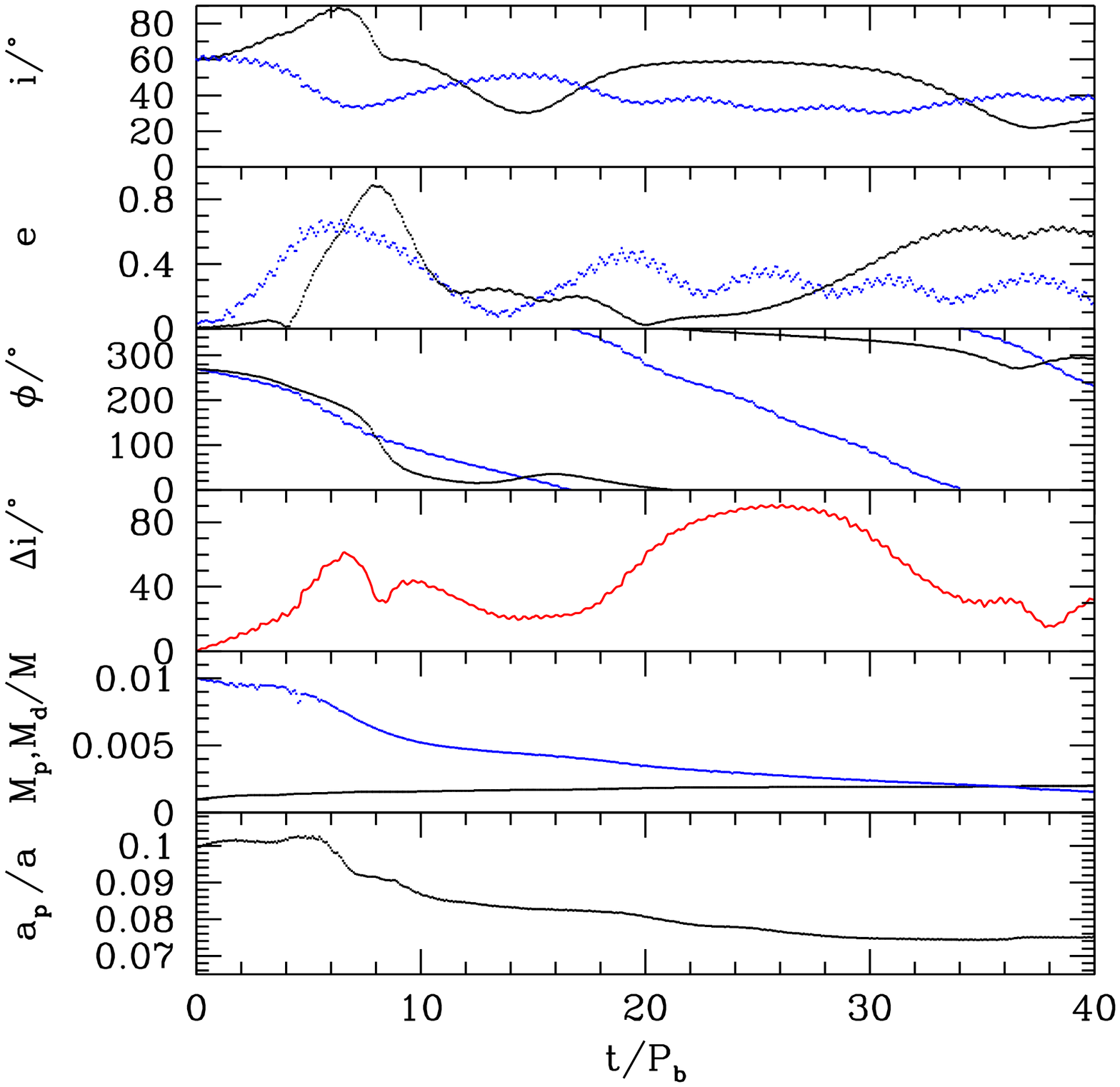}
\includegraphics[width=6.5cm]{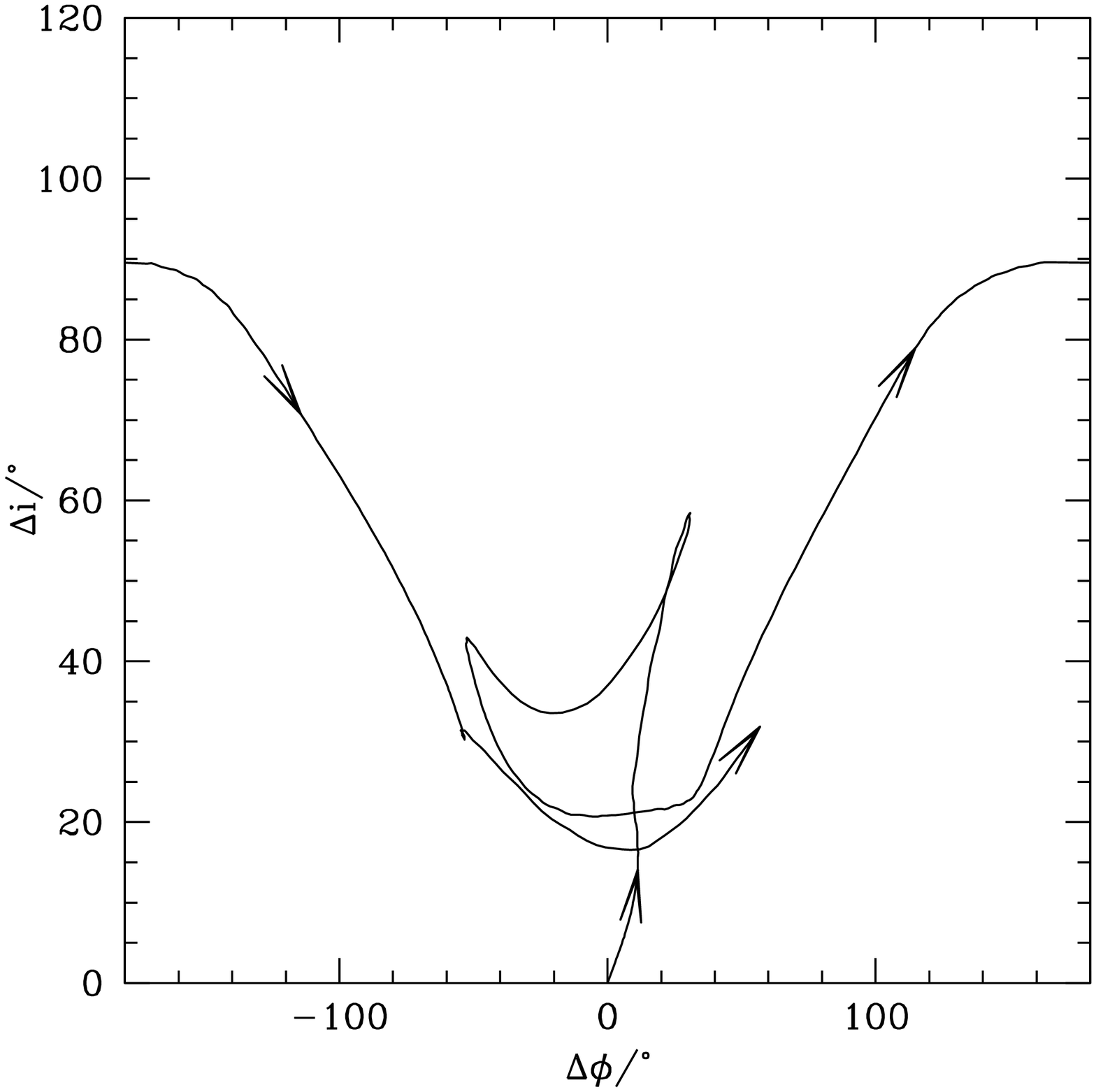}
\end{center}
\caption{Evolution of a planet--disc system around one component of a
  binary. The planet and the disc are initially coplanar 
  but misaligned by $60^\circ$ to the binary orbital plane. The planet
  has an initial mass of $0.001\,M$ and an orbital radius
  of $0.1\,a$ from the primary star. Black lines refer to 
  the planet and blue lines to the disc.  Left panels:
  inclination angle relative to the binary orbital plane (top row),
  eccentricity (second row), phase angle (third row), relative
  planet--disc tilt ($\Delta i$, fourth row), planet and disc masses
  (fifth row) and semi-major axis of the planet (bottom row). Right
  panels: phase portrait of the relative planet--disc tilt versus the
  nodal phase difference, $\Delta \phi=\phi_{\rm p}-\phi_{\rm d}$. Top
  panels: disc of mass $M_{\rm d}=0.001\,M$. Bottom panels: disc of
  mass $M_{\rm d}=0.01\,M$.}
\label{ecc}
\end{figure*}

We first consider a highly misaligned system ($i=60^\circ$) 
for disc masses of $M_{\rm d}=0.001\,M$ and $M_{\rm d}=0.01\,M$. The 
higher disc mass case has previously been studied by \citet{Picogna2015}, 
using similar simulations. We show our results, which are consistent, in order
to establish a benchmark for the lower inclination cases.

Fig.~\ref{ecc} shows the inclinations $i$, eccentricities $e$ and 
phase angles $\phi$ of the disc and planet. The 
relative tilt angle between the planet and the disc is,
\begin{equation}
\Delta i = 
\cos^{-1} ({\bm { l_{\rm p}}  \bm{.l_{\rm d}}}),
\end{equation}
where $\bm  l_{\rm p}$ and $\bm {l_{\rm d}}$ are unit vectors 
of the angular momentum of the planet and
disc respectively. We also plot the mass of the disc, $M_{\rm d}$, the 
mass of the planet, $M_{\rm p}$, and the semi--major axis of the 
planet, $a_{\rm p}$. The disc properties, measured at $r=0.2\,a$, 
are representative of the whole outer disc because the internal 
communication time is short \citep{Martinetal2014b,Fu2015}.
  
In the low disc mass case (upper panels) the planet and disc initially 
execute almost independent tilt oscillations. For a test particle 
in a binary KL oscillations occur on a timescale, 
\begin{equation}
\frac{\tau_{\rm KL}}{P_{\rm b}}\approx \frac{M_1+M_2}{M_2}\frac{P_{\rm b}}{P_{\rm p}},
\label{eq_KL_timescale}
\end{equation}
where $P_{\rm p}$ is the orbital period of the
planet and $P_{\rm b}$ that of the binary
\citep[e.g.][]{Kiseleva1998}. For our parameters we would estimate 
$\tau_{\rm KL} \approx 44\,P_{\rm b}$, similar to the value observed 
numerically, and consistent with no significant disc modification. The 
phase portrait shows that the system is in a circulating phase, where 
the disc and planet precess independently. For the larger disc mass (lower panels) 
there is stronger coupling and this results in stronger misalignment and 
higher eccentricity. Initial libration (with locked precession) transitions 
to circulation as the disc mass decreases.

Fig.~\ref{surfdens} shows the surface density evolution for the low 
mass disc. In this high inclination case, the typically large relative tilt 
between planet and disc allows gas to move inward past the planet, 
and the disc mass drops significantly due to stellar accretion. The accretion 
timescale is roughly coincident with the viscous timescale \citep{Lubow2015b}. 
Planetary accretion is small, but there is migration 
which \cite{Picogna2015} attribute to friction as the planet crosses the 
disc plane.

A key question is whether the oscillations seen in the hydrodynamic simulations 
would persist after disc dispersal. This will happen if the planet has an inclination 
exceeding the critical KL angle when the gas becomes dynamically 
negligible, and will depend on the timescale for dispersal. Observations and 
theoretical models of disc dispersal by photoevaporation suggest that a rough estimate of 
this timescale is $10^5\,\rm yr$ \citep{Wolk1996,Alexander2014}. The dispersal 
time will be short compared to the KL timescale (equation~\ref{eq_KL_timescale}) for  
sufficiently wide binaries, with separations of the order of $10^3 \ {\rm AU}$. In this 
limit the dispersal is effectively instantaneous, and the planets shown in Fig.~\ref{ecc} 
would typically enter KL oscillations (there is a small probability that this would not occur 
in the high mass disc case, as the planet dips below the critical KL angle for two 
brief periods). For closer binaries, a full calculation would need to follow the 
dynamics during the dispersal phase, which might additionally by affected by 
any warping in the system.

\subsection{Critically Misaligned Systems}
\label{crit}

\begin{figure*}
\begin{center}
\includegraphics[width=6.5cm]{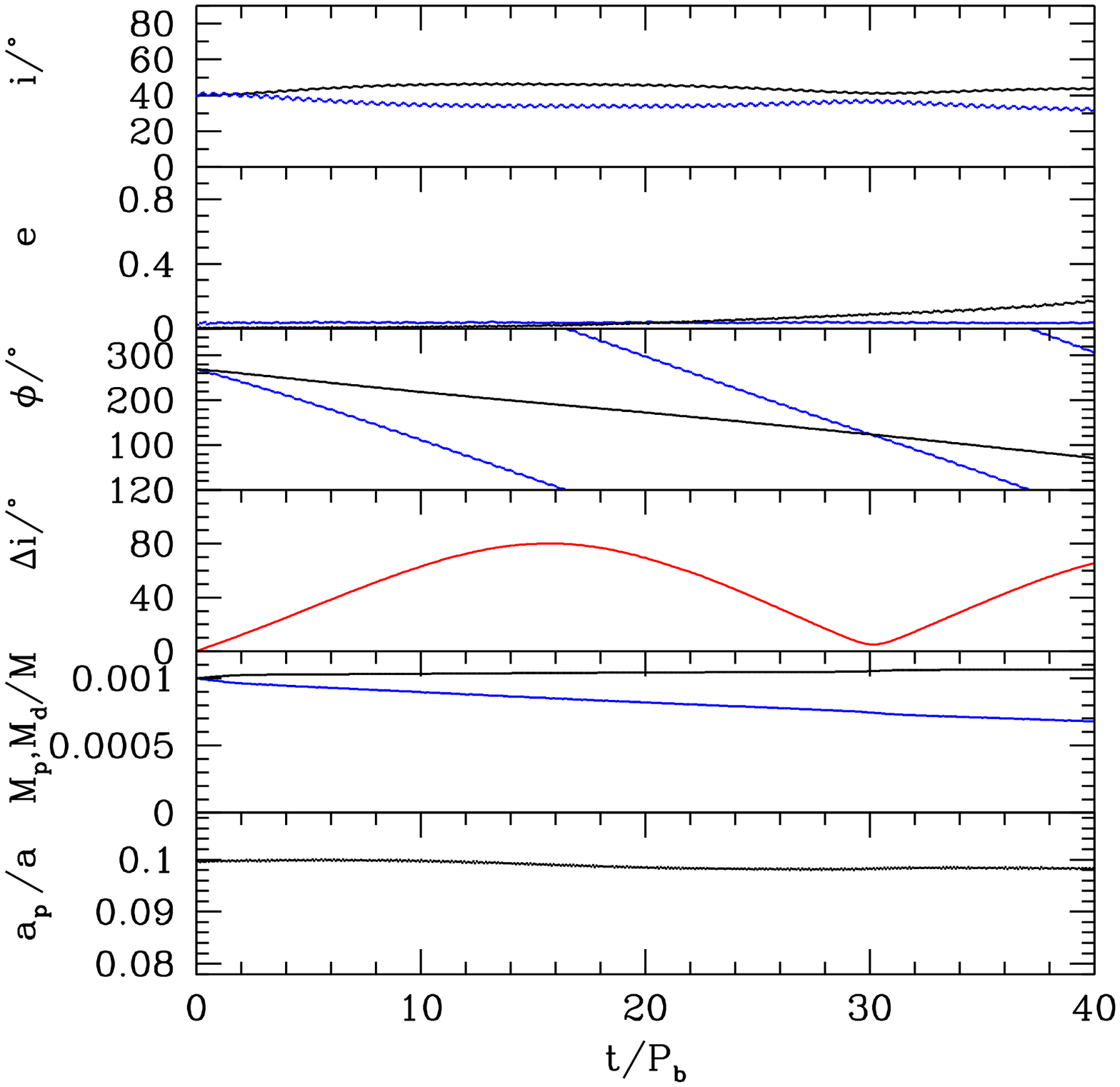}
\includegraphics[width=6.5cm]{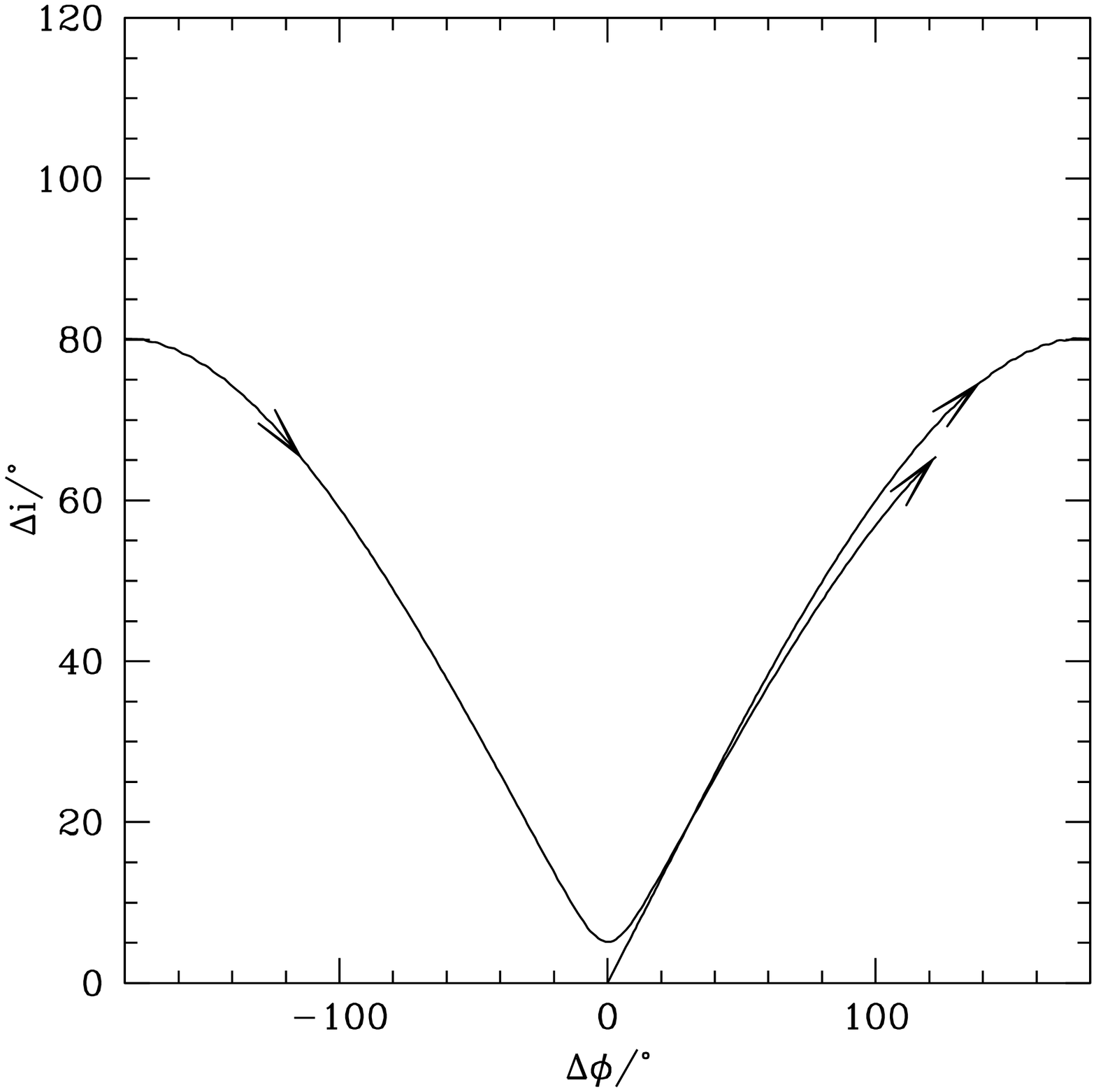}
\includegraphics[width=6.5cm]{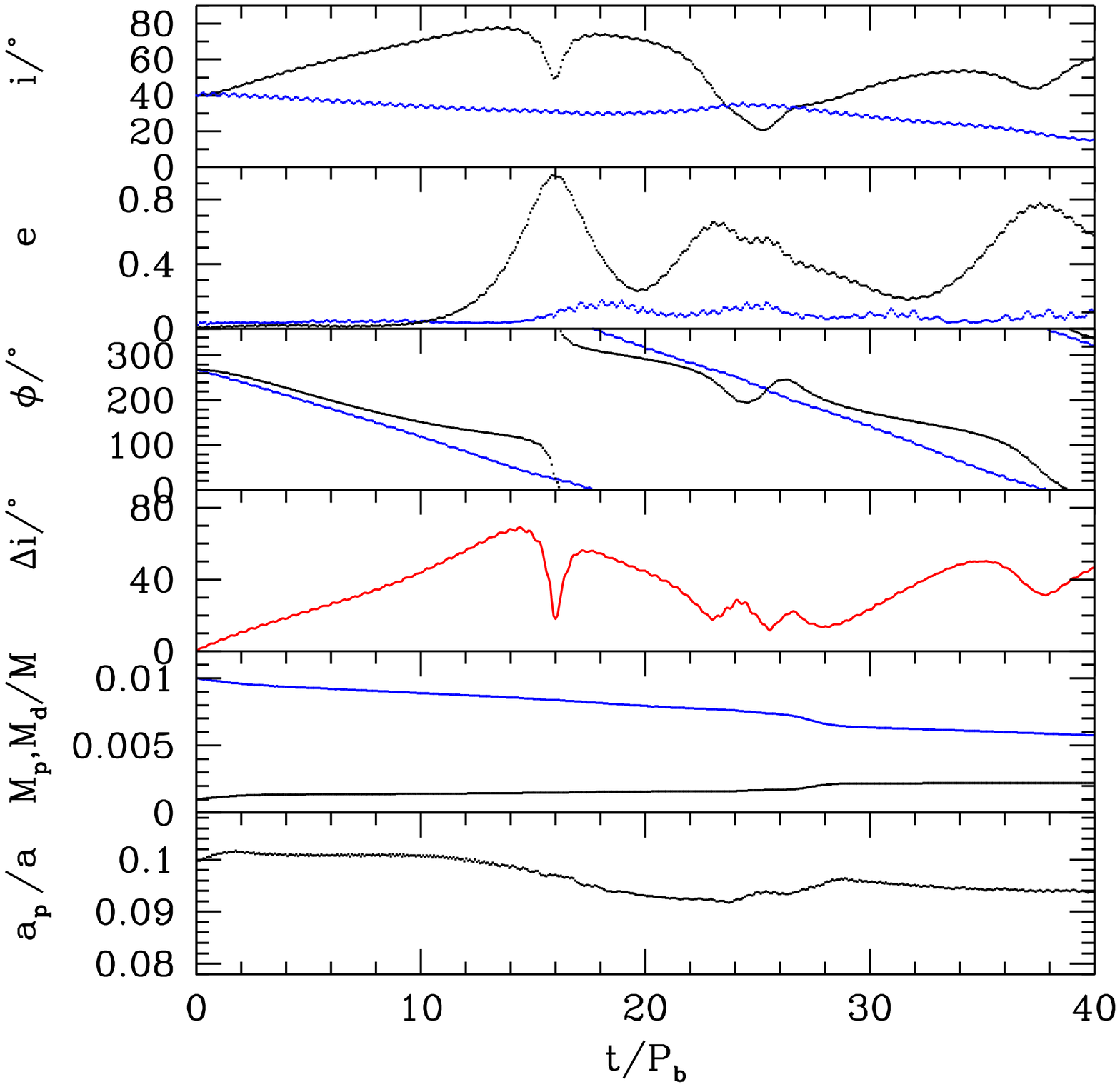}
\includegraphics[width=6.5cm]{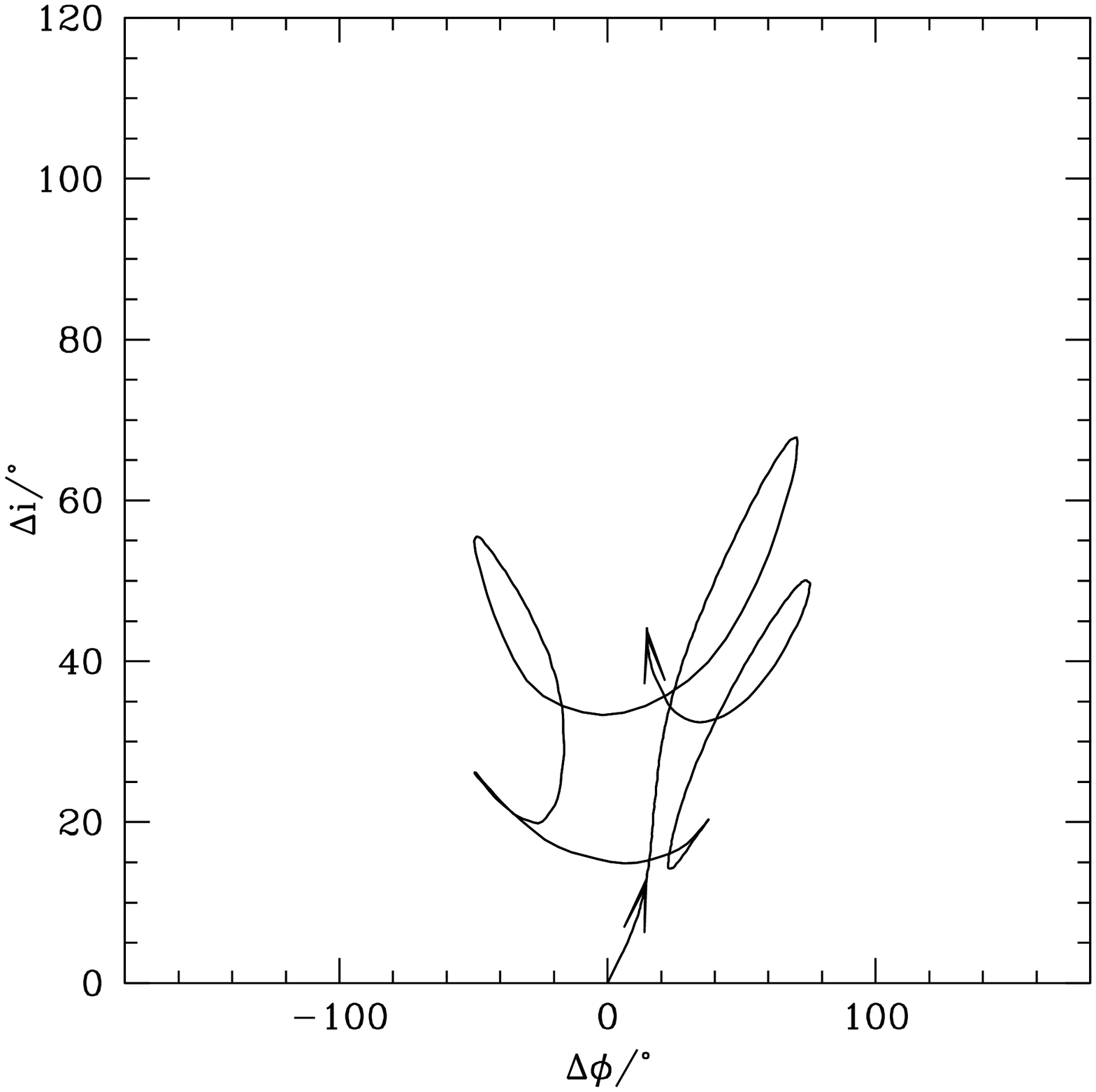}
\end{center}
\caption{Same as Fig.~\ref{ecc} but for initial disc and planet
  inclination of $40^\circ$ to the binary orbital plane. }
\label{ecc3}
\end{figure*}

We now consider planet--disc simulations with an initial inclination of $40^\circ$ to the binary orbital plane, close to the critical angle for KL oscillations.  Fig.~\ref{ecc3} shows the evolution of two systems with disc masses $0.001\,M$ and $0.01\,M$. In both cases, the initial evolution  is the same as the high inclination case. That is, the inclination of the disc decreases, while that of the planet increases. This is the start of a tilt oscillation due to the interaction of the planet and disc, as is discussed further in Section \ref{low}. Thus, even if the planet forms in a circular disc at or slightly below the critical KL angle, the inclination of the planet  increases leading to the likely formation of a KL planet. The disc remains circular and aligns with the binary orbit on a viscous timescale. When the disc mass is high, there is a larger initial increase in planet inclination  and thus stronger planet  KL oscillations. The low mass disc is in a circulating state while the high mass disc is in a librating state  for the length of the simulation.  For the low mass disc case the inclination of the planet is only a few degrees above the critical KL angle, thus the timescale for the KL oscillations is long. At the end of the plotted evolution in the simulation, the eccentricity of the planet is just starting to increase in a KL oscillation. For the larger disc mass, the inclination of the planet is much higher and thus several KL oscillations of the planet are observed during the simulation.

The final outcome of critically misaligned systems is similar to the highly misaligned case. The low mass disc case in Fig.~4 forms a KL planet no matter when the disc is dispersed because the inclination of the planet is always above the critical KL angle. However, the inclination of the planet is only slightly above the critical KL angle and thus the timescale for the KL oscillations is long. For the high mass disc case, the planet displays large KL oscillations during the simulation, but there is a short period of time (at about a time of $t=24\,P_{\rm b}$, lasting for a couple of binary orbits) during which the inclination of the planet is below the critical KL angle. However, the timescale for disc dispersal is longer than this period of time and thus it is most likely that the planet will be a KL planet after the disc disperses. During the time when the inclination is below the critical KL angle, the eccentricity of the planet is high. Thus if the disc was dispersed during this time, the planet would still  have a large eccentricity.

\subsection{Slightly Misaligned Systems}
\label{low}

We now consider conditions under which a system that begins below the critical KL angle can produce a KL planet.  In \cite{Lubow2015b} we found that 
initially mutually coplanar planet--disc systems that are slightly inclined to the
binary orbital plane undergo secular
misalignment tilt oscillations.
For a small mass disc, the planet and disc undergo relative tilt oscillations,
as a consequence of their different nodal precession rates that are due to the binary. However, their tilts
relative to the binary orbital plane are nearly constant. In particular,
the tilt of the planet orbit relative to the binary orbital plane is nearly constant
in time, for sufficiently small disc mass.

For more massive discs, as a consequence of the planet's gravitational interaction with
the disc, the planet's orbital tilt relative to the binary orbital plane can increase \citep[see Fig.~1 of][]{Lubow2015b}. In particular,
in the absence of dissipation, during secular oscillations, the planet's orbital tilt relative to the binary is always greater
than the disc's. This difference in tilt behavior between the planet and disc is a 
consequence of the lower angular momentum of the planet compared to the disc.
The planet's orbital tilt increase is aided by the effects of a secular resonance caused by the disc and binary companion.
The planet's orbital tilt relative to the binary can reach a value of a few times the initial planet--binary tilt for disc
masses that are roughly 5 to 10 times the planet mass. 
As we show below, this boost in planet orbital tilt can
cause a planet whose tilt is below the KL angle to exceed
that angle and form a KL planet.

Another effect is that if the planet accretes mass and momentum from a disc
whose tilt has decayed by dissipation to a coplanar configuration with the binary, the planet
orbit will evolve towards coplanarity with the binary. This effect plays a role
for massive discs.

For  a planet--disc system that begins below the critical KL angle, the tilt oscillations may increase the planet orbit inclination to above $40^\circ$ and form a KL planet. If the disc mass is low, then the amplitude of the tilt oscillations is small and a KL planet cannot be formed. On the other hand, if the disc mass is high, the amplitude of the tilt oscillations is large, but they may be damped by accretion from the disc on to the planet. Thus, the disc mass must be within a certain range for low inclination systems to produce a KL planet. We consider here an inclination of $20^\circ$, but note that the range of disc masses that allow the formation of a KL planet is wider for higher initial inclination systems. 

\begin{figure*}
\begin{center}
\includegraphics[width=6.5cm]{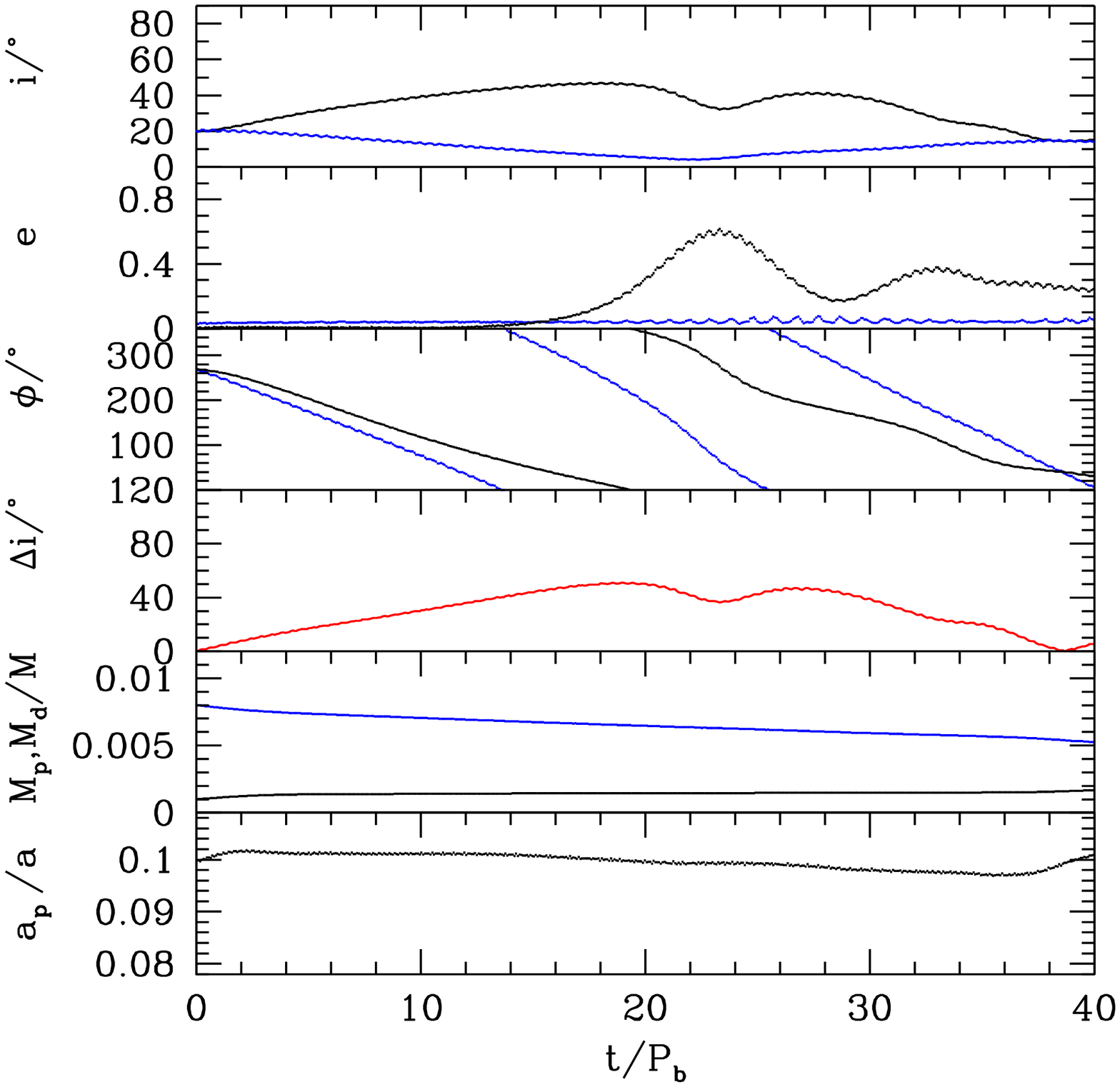}
\includegraphics[width=6.5cm]{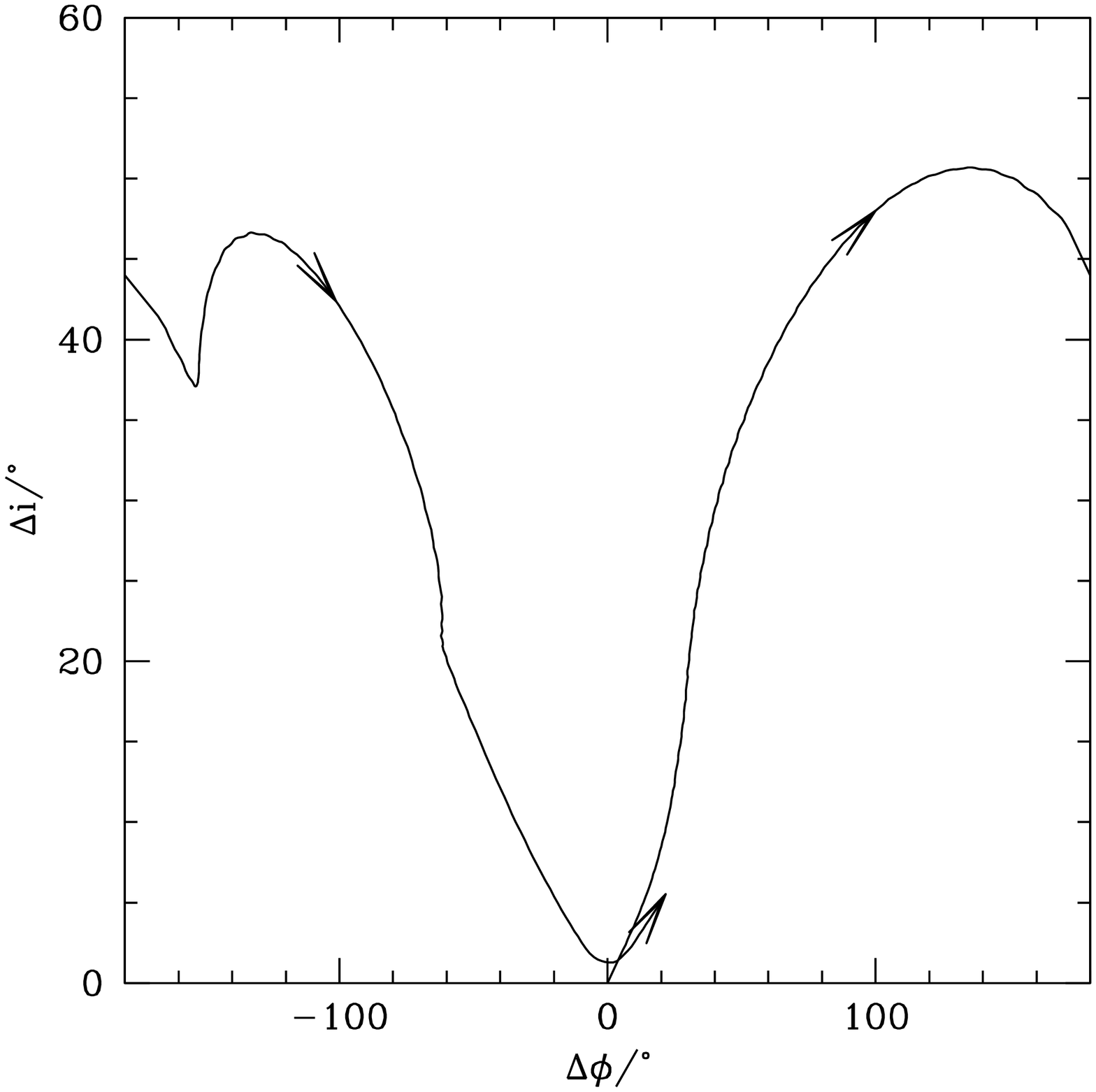}
\end{center}
\caption{Same as Fig.~\ref{ecc} but for initial disc and planet
  inclination of $20^\circ$ to the binary orbital plane. The disc mass
  is $0.008\, M$.}
\label{ecc4}
\end{figure*}

We first consider a disc mass of $0.008\,M$ and show the evolution of the system in Fig.~\ref{ecc4}. The inclination of the planet initially increases during a tilt oscillation until it becomes a KL planet. Eccentricity and inclination oscillations of the planet occur in the simulations. The system is in a circulating phase state. At the end of the simulation, the planet and the disc have the same inclination. This state occurs at the end of one full tilt oscillation. If the simulation were to continue, then we expect further tilt oscillations to occur. Whether or not the system forms a KL planet may be sensitive to the time of the disc dispersal because the planet spends more than half of its time below the critical KL angle. However, during this time, the planet orbit is fairly eccentric. 

We describe here some further simulations that are not shown in the figures. For a slightly larger disc mass of $M_{\rm d}=0.01\,M$ and the same initial inclinations, the disc and the planet are librating, i.e., their phase angles precess together.  However, because there is some damping of the tilt oscillations, the inclination of the planet  only reaches slightly above the critical KL angle (maximum at $i=40.1^\circ$) and thus the eccentricity growth is relatively small (up to about $e=0.2$). For a higher disc mass ($M_{\rm d}>0.01\,M$), the tilt oscillations are more heavily damped by the accretion of disc material on to the planet and the planet does not show KL oscillations. For a lower disc mass of $M_{\rm
  d}=0.005\,M$, the tilt oscillations are not significantly damped, but the inclination of the planet is only just over the critical KL angle at its maximum of $i=40.8^\circ$. Thus, the planet undergoes KL oscillations, but the eccentricity growth is small (up to $e\approx 0.1$). For a smaller disc mass ($M_{\rm d}<0.005\,M$), the tilt oscillations are not large enough to form a KL planet. Thus, for the parameters chosen and initial inclination of $20^\circ$, there is a range of initial disc masses that can produce a KL planet, $0.005\, M \lesssim M_{\rm d} \lesssim
0.01\, M$.  During the simulation, the planet spends only some time with an inclination above the critical KL angle. In all cases, whether the planet will be a KL planet after the disc dispersal depends on the timing.  Although, the probability of forming a moderately eccentric planet ($e\gtrsim0.2$) is high.

\section{Discussion}
\label{discussion}

\begin{figure*}
\begin{center}
\includegraphics[width=20cm]{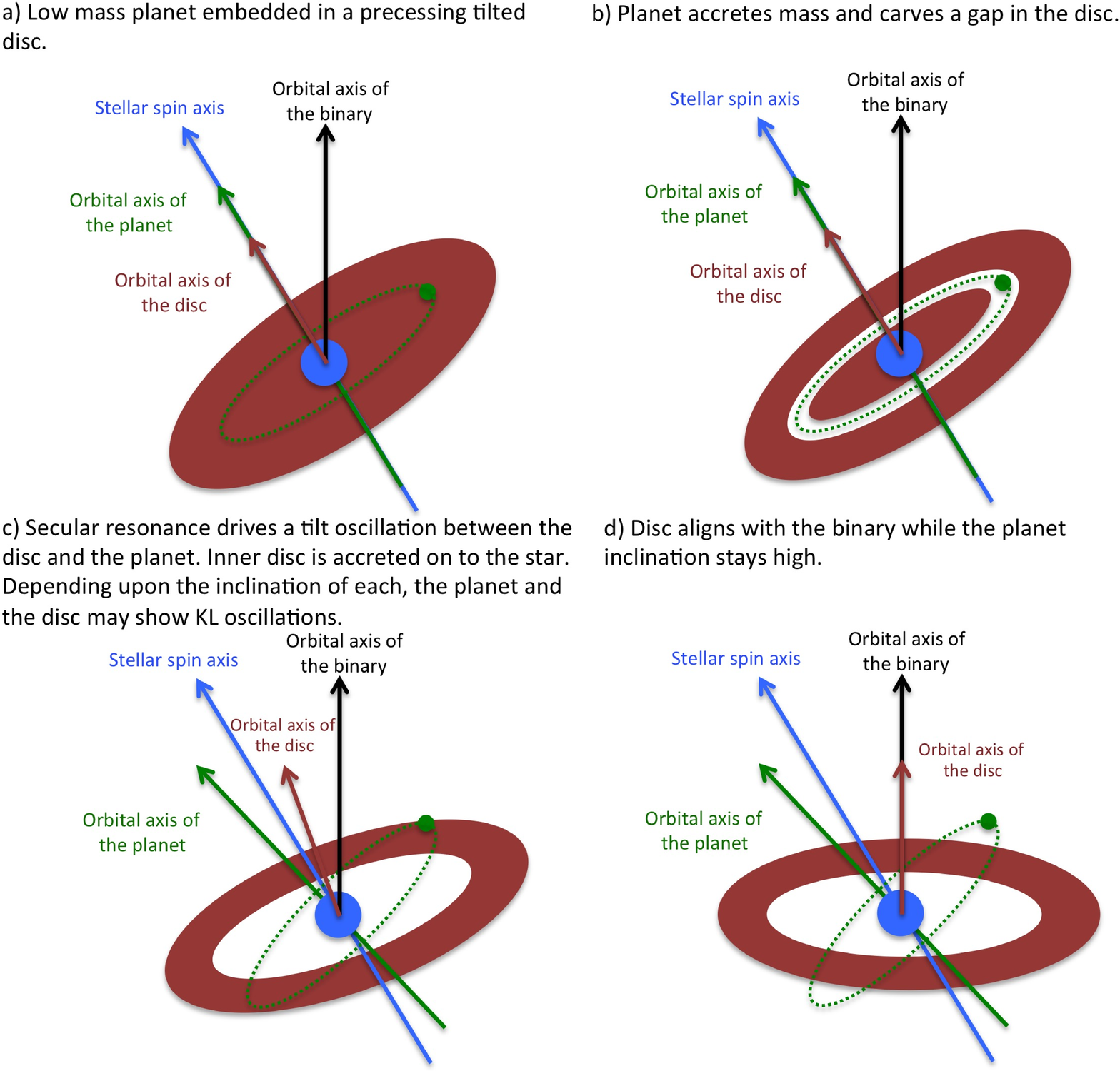}
\end{center}
\caption{Sketch of the steps that lead to the formation of a KL planet in a tilted disc in a binary system. Top left: A low mass planet forms in a disc that is misaligned to the binary orbital axis. If the disc is above the critical KL angle, then self--gravity must be strong enough to prevent KL oscillations of the disc. If the disc is below the critical KL angle then the disc need not be self--gravitating to remain circular. Initially the disc, planet and stellar spin are aligned. However, the disc and the planet together are precessing about the orbital axis of the binary. Top Right: As the planet accretes material from the disc, it becomes massive enough to open a gap in the disc. Bottom left: The secular resonance is now strong enough for a tilt oscillation to begin. The inclination of the planet to the binary orbital axis increases while that of the disc decreases. The planet and the disc decouple, although their precession angles may be in phase (the librating state) or not (circulating state). If the planet is above the critical KL angle, it shows KL oscillations. Furthermore, the disc may also be above the critical KL angle and showing damped KL oscillations. Bottom Right: On a longer timescale the disc aligns with the binary orbital plane (according to Fig.~\ref{klalign}), while the planet can remain above the critical KL angle. }
\label{cartoon}
\end{figure*}

The disc around one component of a binary may initially be aligned with the spin of the star. However, the precession induced by the binary causes the disc and the stellar spin to become misaligned.  \cite{Batygin2012} showed that a misaligned precessing self--gravitating disc around one component of a binary leads to the orbits of planets being misaligned with the spin of the star if the planets remain coplanar to the disc. While we find that the planet and the disc most likely do not remain locked together, the outcome of a misaligned planet is likely.   In Fig.~\ref{cartoon} we sketch the evolution of misaligned planet--disc systems that lead to the formation of a KL planet. 

The upper left panel of Fig.~\ref{cartoon} depicts a low mass planet forming in a misaligned disc around one component of a binary system.  A disc undergoing KL oscillations is 
generally noncircular and
would be subject to destructive impacts among newly-formed planetesimals \citep[e.g.][]{Marzarietal2009,Fragneretal2011}. Therefore, KL oscillations would be expected to inhibit planet formation in the standard core accretion model. If the disc tilt is above the critical KL angle, the disc must be self--gravitating in order to avoid KL oscillations during the early times and allow planet formation to take place. However, if the disc is not initially above the critical KL angle, then it does not need to be self--gravitating to remain circular and form a planet. In both cases, the disc and low mass planet precess together about the binary orbital axis.

In a subsequent stage, shown in the top right panel of Fig.~\ref{cartoon}, the planet accretes material from the disc and becomes large enough to clear a gap in the disc. The inner disc becomes disconnected from the outer disc and it accretes on to the central star on a short timescale. Because of the gap between the planet and the disc, the torque from the binary on to the planet may dominate the torque from the disc on to the planet. In the lower left panel, a secular resonance drives a tilt oscillation. Initially, the inclination of the disc moves towards the binary orbital plane, while the planet's inclination increases. The disc and the planet become decoupled. Depending on the mass of the disc, the planet and the disc may have the same precession rate (the librating state) or they may precess individually (the circulating state). If the inclination of the planet is above the critical KL angle, then KL oscillations take place. Similarly for the disc, if it is above the critical KL angle and self--gravity is not sufficiently strong, then it will display damped KL oscillations. While the disc and the planet are decoupled, the disc spreads inwards past the orbital radius of the planet. The lower right panel shows that on a longer timescale the disc aligns with the binary orbital plane, according to the description in the introduction and Fig.~\ref{klalign}, while the inclination of the planet may remain high. This cartoon depicts the inclination angles, but not the precession angles for the disc and planet.

\section{Conclusions}
\label{conc}

The dynamical and tidal evolution of KL planets provide a potential explanation for several observed properties of massive extrasolar planetary systems. Once the gas disc has been dispersed, both the conditions needed for KL oscillations to develop, and their outcome, are well-established. In this paper we have argued that the {\em prior} phase of evolution, when the gas disc from which the planet formed is present and dynamically important, can substantially modify the range of systems that ultimately form KL planets. The new elements that we consider involve three coupled effects that are summarised below. 1) Tilt oscillations occur between the planet and the disc, involving the binary. They are the result of a four body effect. For this type of interaction, the disc (planet) tilt changes relative to the binary orbital plane only if the planet (disc) is present. These tilt oscillations are present even at small tilt angles \citep{Lubow2015b}. 
2) Global KL tilt and eccentricity oscillations occur in the disc due to driving by the binary
\citep{Martinetal2014a}. They are the result of a three body effect.  These disc oscillations occur even if no planet is present. KL oscillations take place only for substantial misalignment ($> 40 ^\circ$) between the disc and binary orbital planes.  3) There is a long-term evolution of the disc plane that is driven through the viscous damping of a warp in the disc that is caused by the binary \citep[e.g.,][]{Kingetal2013}. The relative strength of these effects depends upon the planet-disc mass ratio, on the importance of self-gravity within the disc, and on the magnitude of the disc viscosity.

We have used hydrodynamic simulations to study the long term evolution of a planet that forms in a protoplanetary disc that is initially misaligned to the orbital plane of a binary companion star. For the specific parameters that we have focused on, we find that the hydrodynamic evolution of the coupled planet-disc systems {\em broadens} the range of initial systems that form KL planets. Depending upon the mass of the disc, we find that KL planets could form from systems in which the initial misalignment angle lies in the range $20^\circ \lesssim i \lesssim 160^\circ$. The formation of a KL planet also depends on both the speed of disc dissipation, and on the specific time during the oscillation cycle that the disc is lost, and is thus somewhat stochastic. However, even if the disc disperses while the planet is below the critical KL angle, a moderately eccentric planet is expected.

\section*{Acknowledgments} 
SHL acknowledges support from NASA grant NNX11AK61G.  CN was supported by the Science and Technology Facilities Council (grant number ST/M005917/1). Research in theoretical astrophysics at Leicester is supported by an STFC Consolidated Grant. PJA acknowledges support from NASA through grant NNX13AI58G  and from the NSF through grant AST 1313021.  We acknowledge the use of SPLASH \citep{Price2007} for the rendering of the figures. Computer support was provided by UNLV's National Supercomputing Center.

\bibliographystyle{mn2e} 
\small

\label{lastpage}
\end{document}